\documentclass[11pt]{article}
\usepackage[margin=1.9cm]{geometry}
\usepackage[T1]{fontenc}
\usepackage{stix2}                 
\usepackage{graphicx}
\usepackage{booktabs}
\usepackage{colortbl}
\usepackage{xcolor}
\usepackage{amsmath}
\usepackage{amssymb}
\usepackage{anyfontsize}
\usepackage[numbers,super,sort&compress]{natbib}  
\usepackage{caption}
\usepackage{hyperref}
\hypersetup{colorlinks=true,allcolors=blue}
\graphicspath{{figures/}}
\begin{document}

\title{\bfseries Agentic Autoresearch for CT Reconstruction}
\author{%
\begin{minipage}{0.95\textwidth}\centering
\normalsize Andreas Maier\textsuperscript{1,5}, Lucas Kachelrie\ss\textsuperscript{1,2}, Siming Bayer\textsuperscript{1}, Yixing Huang\textsuperscript{3}, Yan Xia\textsuperscript{2}, Amber Simpson\textsuperscript{4}, Moritz Zaiss\textsuperscript{2,5}\\[0.9em]
\small
\textsuperscript{1}Pattern Recognition Lab, Friedrich-Alexander-Universit\"at Erlangen-N\"urnberg (FAU),  Erlangen, Germany \\
\textsuperscript{2}Universit\"atsklinikum Erlangen (UKER),  Erlangen, Germany \\
\textsuperscript{3}Peking University,  Beijing, China \\
\textsuperscript{4}University of Alberta,  Edmonton, Canada \\
\textsuperscript{5}Department Artificial Intelligence in Biomedical Engineering (AIBE), Friedrich-Alexander-Universit\"at Erlangen-N\"urnberg (FAU),  Erlangen, Germany\\[0.6em]
Correspondence: \texttt{andreas.maier@fau.de}
\end{minipage}}
\date{}
\maketitle

\begin{abstract}\noindent
\textbf{Background.} Deep learning has produced a large, diverse body of CT reconstruction methods, but fair comparison remains labor-intensive and largely manual, and many benchmarks rely on idealized data.
\textbf{Purpose.} We ask whether a large language model (LLM) agent can do the labor of reconstruction research on its own, and whether a ranking measured on ideal data predicts how methods behave under realistic noise.
\textbf{Methods.} We built an agentic loop: the agent edits a solver, runs a short cluster job, reads one frozen metric, and revises. The metric is a calibrated headroom score, $\mathrm{hr}=\max(0,\,1-\mathrm{RMSE}/\mathrm{RMSE}_{\mathrm{FBP}})$, computed inside the field of view, and every method is grounded in the same differentiable fan-beam projector. We benchmarked 26 methods on two problems, Mayo low-dose CT (noise-limited) and a 128-view sparse-view breast task from the noiseless DL-Sparse-View Challenge, selecting each method's best iteration on validation and reporting its held-out test score (per-case mean $\pm$ SD). To inspect distribution shift, every trained Breast CT model was then re-scored on noisy breast inputs ($I_0=10^5$ photons) \emph{without retraining}, and separately retrained on matched-noise data.
\textbf{Results.} The agentic loop independently implemented, tuned, and benchmarked all 26 state-of-the-art methods, and recombined them into a compact solver of 969 parameters that ties the top Mayo tier at the 1\,\% level with 0.4\,\% of the champion's parameters; its best architecture is problem-dependent: a denoiser for noisy Mayo, a filtered data-consistency and primal-dual unroll for sparse-view breast (195 parameters). Benchmarking gives a small tier of statistically indistinguishable top methods, not one winner. Mild input noise nearly inverts the breast ranking: the noiseless champion (supervised image denoiser, hr 0.89) collapses to hr 0.00, while a learned primal-dual method rises from mid-pack to champion (0.72 $\to$ 0.93). Physics-regularized and hand-crafted-smoothing methods rise; supervised image-domain denoisers collapse. Retraining restored the ranking partly.
\textbf{Conclusions.} An LLM agent can implement, tune, and benchmark reconstruction methods at scale under a fixed metric and budget, though it invented no new methods and needed human guidance for the most parameter-efficient ones. More importantly, an ideal-data leaderboard does not predict robustness: it can invert under a small, realistic perturbation. The inversion is a transfer effect, not a permanent deficit; retraining on matched noise restores much of the clean ranking (Spearman $\rho$ from $0.04$ to $0.61$). But noise is only the easiest confounder in an open-ended, shifting set (beam hardening, scatter, anatomy, disease), so no single-factor challenge can certify generality. Benchmarks should model a broad spectrum of realistic factors at once. Agentic autoresearch is an enabler to scale experiments thereon and to implement fair comparisons.%
\end{abstract}
\vspace{0.5em}\noindent\textbf{Keywords:} CT reconstruction | sparse-view | low-dose | known operators | deep learning | benchmarking | large language model agent | noise robustness

\section{Introduction}\label{sec:intro}

Deep learning has changed CT reconstruction. A decade of work has produced many learned reconstructors. Examples are unrolled iterative networks, learned primal-dual schemes, variational networks\footnote{A variational network (Hammernik et al.) is an unrolled network whose learned regularizer is a filter-bank ``field of experts''.}, image-domain and dual-domain denoisers, diffusion priors (generative models, unrelated to physical diffusion), and per-scene implicit representations.\cite{sidky2022,adler2018lpd,hammernik2018vn,wurfl2016deep,wuerfl2018,ongie2020} Each promises a good image from fewer views or a lower dose than filtered back-projection (FBP). The best of them deliver. At first glance the research program looks simple: pick the strongest network, train it, reconstruct.

Yet, a purely data-driven reconstructor learns a map that is not tied to the physics. Such a map is free to \emph{hallucinate}.\cite{huang2018some} It can add anatomy that the measurements do not support.\cite{maier2019gentle} This is not a small tuning issue. It is a structural risk. The field's durable answer is to put the physics back in. We embed the \textbf{known operator} --- the differentiable forward and back projector --- into the learned pipeline.\cite{maier2019knownop,maier2022review} In fact, this trade is quantified. Constraining a network with an exact operator provably cannot raise, and usually lowers, the maximum error bound. In practice it also lowers the number of parameters and the amount of training data needed.\cite{maier2019knownop} A recent deep risk estimator establishes this benefit on theoretical grounds.\cite{maier2026deep}

There is a second problem, and it sits one level up. The body of methods is itself the product of slow, largely manual work. The reconstruction problem can be framed as a learned inverse problem,\cite{ongie2020} but to compare or improve any method, a researcher must implement a solver, wire it to the data and geometry, launch a job, read a metric, form a hypothesis, change one thing, and repeat. This takes days, weeks, sometimes even months per method. Recent ``autoresearch'' ideas imagine automating this loop, an automated research cycle in which the software itself proposes a change, runs the experiment, reads the result, and iterates.\cite{karpathy2026autoresearch,lu2024aiscientist} Large language model (LLM) agents make this practical. An LLM is a neural network trained to predict text; an \emph{agent} wraps such a model in a loop so that it can act (read files, write code, launch jobs) rather than only answer questions, and a loop driven this way we call \emph{agentic}. General-purpose coding agents already resolve real software-engineering tasks,\cite{yang2024sweagent,jimenez2024swebench,maier2026vibecoding} and physics-informed agentic development has begun in adjacent imaging fields.\cite{zaiss2026agent4mr} \emph{Can an LLM agent do reconstruction research?} Not just tune one number, but implement an unfamiliar method, diagnose why a reconstruction fails, edit its own solver code, and benchmark the result fairly, if we ground it in the same two things that would guide a human: a differentiable projector and a fair, frozen metric.

We built such a loop and let an agent run it. With it, we benchmarked 26 methods across two CT problems. One is noise-limited (Mayo low-dose\cite{mccollough2017mayo}); the other is incompleteness-limited (128-view sparse-view breast, from the noiseless DL-Sparse-View Challenge\cite{sidky2022}). We then asked a second question that the ideal-data benchmark cannot answer on its own: does the ranking survive a little noise? To answer it, we added mild Poisson noise to the test inputs and re-scored every trained model without retraining, and then retrained the models on matched noise to see whether the ranking returns.

This paper makes five contributions. (i) An agentic CT-reconstruction loop grounded in a differentiable projector and a frozen calibrated metric, with full provenance. (ii) An autonomous implementation and fair benchmark of 26 methods across two regimes. (iii) An evidence-derived compact solver, competitive with the top tier at $\sim$1--2\,\% of its parameters, whose optimal form is problem-dependent. (iv) A robustness result (ideal-data rank does not predict, and can invert under, noise) and the practical lesson it carries for how we build benchmarks. (v) We release all 26 methods, our compact solver, and the agentic loop as open source (\url{https://github.com/akmaier/Agent4CT}).

\section{Materials and Methods}\label{sec:methods}

\subsection{The agentic loop}\label{sec:loop}

The agent is driven by a single large language model, Claude Opus~4.8 (Anthropic; model \texttt{claude-opus-4-8}, 1M-token context). It runs as an autonomous coding agent with standard file, shell, and version-control tools and access to a compute cluster; no external tools or plugins beyond these were connected. The loop itself is the one popularized by Karpathy's \emph{autoresearch}:\cite{karpathy2026autoresearch} the software proposes a change, runs the experiment, reads the metric, and keeps or discards. Its behaviour was shaped by three fixed inputs: a written solver recipe and repository guide, the frozen validation metric, and the per-iteration loop contract that follows. The agent works in a fixed cycle. First it reads the previous result and names the failure. By \emph{failure} we mean the specific way the current best reconstruction falls short of the frozen metric: a stalled or low headroom score, or a dominant error mode read off the comparison figure (residual streaks, over-smoothing, noise amplification). It then changes one thing, a hyper-parameter or the solver code itself, and states a hypothesis. A single short cluster job runs under a fixed compute budget. The agent reads the frozen metric and accepts or discards the change. The cycle then repeats.

\begin{figure}[t]
\centerline{\includegraphics[width=0.62\linewidth]{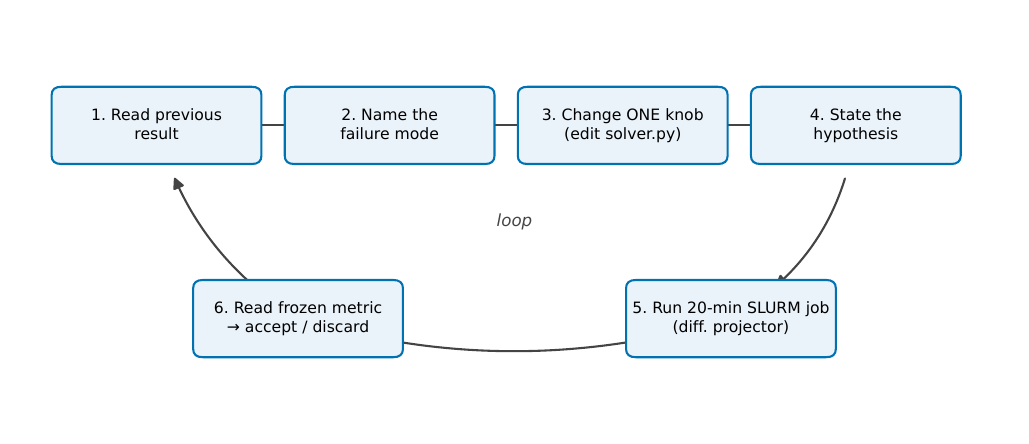}}
\caption{The agentic autoresearch loop and its provenance. The agent reads the previous result, names the failure, changes one thing, states a hypothesis, runs one short cluster job grounded in the same differentiable fan-beam projector, reads the frozen calibrated-headroom metric, and accepts or discards the change. Every iteration writes an immutable record (configuration, reconstruction, metric, comparison figure).\label{fig:loop}}
\end{figure}

Every run is grounded in the same \emph{differentiable} fan-beam projector (PYRO-NN\cite{syben2019pyronn}), a forward/back-projector built so that gradients flow through it and can therefore sit inside a trainable network optimized by back-propagation. Each run also writes an immutable record: the configuration, the reconstruction, the metric, and a comparison figure. We enforced a fixed wall-clock budget per iteration (20~min on breast, matched on Mayo) so that all methods are compared under equal compute. A human coordinator supervised the loop, setting targets, redirecting across problems, and auditing for padding and provenance gaps (Section~\ref{sec:discussion}). The loop is sketched in Figure~\ref{fig:loop}.

\subsection{A single framework for 26 methods}\label{sec:framework}

We describe the forward model first. For the reconstruction, the scanner is assumed as a linear operator. The sinogram is the set of line integrals of the image:
\begin{equation}\label{eq:forward}
\mathbf{g} = \mathbf{A}\,\mathbf{x} + \boldsymbol{\varepsilon}.
\end{equation}
Here $\mathbf{x}$ is the image, $\mathbf{A}$ the discrete fan-beam projector, $\mathbf{g}$ the measured projections, and $\boldsymbol{\varepsilon}$ the noise. For the noiseless breast data $\boldsymbol{\varepsilon}=\mathbf{0}$. Reconstruction means recovering $\mathbf{x}$ from $\mathbf{g}$. With 128 views, $\mathbf{A}$ is under-determined.

Almost every method is one instance of the same regularized inversion. We trade fit to the measurements against a prior:
\begin{equation}\label{eq:variational}
\hat{\mathbf{x}}=\arg\min_{\mathbf{x}}\ \tfrac12\|\mathbf{A}\mathbf{x}-\mathbf{g}\|_2^2+\lambda\,\mathcal{R}(\mathbf{x}).
\end{equation}
A method is fixed by two choices. The first is the \textbf{data-consistency operator} $\mathbf{D}$, which maps the measurement residual back to image space and so pulls the image toward agreement with the sinogram. The second is the \textbf{prior} $\mathcal{R}$ (also called the regularizer): it encodes what a plausible image looks like and sets the fit-versus-smoothness trade-off. Many solvers approximate Eq.~(\ref{eq:variational}) by an \emph{unrolled} proximal-gradient scheme of $K$ steps. Unrolling fixes the iteration to $K$ steps and makes each step a trainable network layer, so the whole fixed-length scheme is trained end-to-end:
\begin{equation}\label{eq:unroll}
\mathbf{x}_{k+1} = \mathbf{x}_k - \alpha_k\,\mathbf{D}\big(\mathbf{A}\mathbf{x}_k-\mathbf{g}\big) + \mathcal{R}_\theta(\mathbf{x}_k).
\end{equation}
Each step is a proximal-gradient update: a gradient step toward data fit through $\mathbf{D}$, then a ``proximal'' step --- the learned prior $\mathcal{R}_\theta$ --- that cleans the image. The form of $\mathbf{D}$ matters. Here, the raw adjoint $\mathbf{D}=\mathbf{A}^{\!\top}$ back-projects the residual and re-injects sparse-view streaks, whereas the filtered map $\mathbf{D}=\mathrm{FBP}(\cdot)$ (a ramp/Hann-filtered back-projection) does not. This distinction is the key lever on the sparse-view problem.

Some methods add a learned dual in the measurement domain. This is the learned primal-dual (LPD) block.\cite{adler2018lpd} It learns updates in both the image domain (the \emph{primal}) and the measurement/sinogram domain (the \emph{dual}). A sinogram-space memory $\mathbf{h}$ and the image $\mathbf{x}$ update together:
\begin{equation}\label{eq:lpd}
\mathbf{h}_{k+1}=\Gamma_\phi\!\big(\mathbf{h}_k,\ \mathbf{A}\mathbf{x}_k-\mathbf{g}\big),\qquad \mathbf{x}_{k+1}=\Lambda_\theta\!\big(\mathbf{x}_k,\ \mathbf{D}(\mathbf{h}_{k+1})\big).
\end{equation}
The convolutions $\Gamma_\phi,\Lambda_\theta$ are small and weight-tied. We zero-initialize their final layer, so the block is the identity at the start and cannot regress.

Table~\ref{tab:taxonomy} places all 26 established methods on two axes. Axis~A is physics engagement: how much $\mathbf{A}$ is used at inference (none, a single data-consistency step, an in-loop unrolled scheme, or a full per-scene fit). Axis~B is the prior source, and its categories are worth naming plainly. \emph{Hand-crafted} priors (total variation, bilateral) encode smoothness by hand. \emph{Supervised} priors are trained on paired examples with a known answer (low-dose input, full-dose truth). \emph{Self-supervised} priors train without a clean reference, using structure in the data itself; Noise2Inverse (N2I), for example, splits the measured projections into subsets so one subset's reconstruction predicts another's. \emph{Generative} priors are diffusion models --- networks that make realistic images by reversing a gradual noising process --- used to ``imagine'' plausible detail. \emph{Implicit/per-scene} methods fit a small network to each single scan that returns the attenuation at any queried coordinate, rather than storing a pixel grid; neural fields and Gaussian splatting (which models the volume as many small 3-D Gaussian blobs) are examples. A \emph{foundation model} is a large network pre-trained once on generic data and applied here with no task-specific training (``zero-shot''); we include the Reconstruct Anything Model.\cite{terris2025ram} This table also names the agent's job in the compact study: re-select the pair $(\mathbf{D},\mathcal{R})$. Full per-solver configurations are in Supplement~S1.

\textbf{Search protocol.} Each of the 26 established methods was tuned by 20 autoresearch iterations of the loop in Section~\ref{sec:loop}. The 27th method, our compact solver, was searched for 40 iterations with the explicit aim of an ultra-low-parameter recombination of the pair $(\mathbf{D},\mathcal{R})$. The parameter budget was set by instruction, to minimise parameter count while holding headroom, not as a term in the frozen metric; Figure~\ref{fig:params} shows the resulting accuracy/parameter front. During its first 20 iterations the agent worked alone and was free to recombine any component of the other 26 solvers. During the second 20, a human operator added mild suggestions about directions not yet tried. This was the only stage where the human shaped the search beyond setting targets and auditing.

\begin{table}[t]
\caption{Solver taxonomy (condensed). Each family is placed on Axis~A (physics engagement, i.e.\ how the operator $\mathbf{A}$ is used at inference) and Axis~B (prior source $\mathcal{R}$), with a source citation per family. The full per-solver configuration and parameter counts are in Supplement~S1.\label{tab:taxonomy}}
\centering
\setlength{\tabcolsep}{4pt}
\resizebox{\columnwidth}{!}{%
\begin{tabular}{@{}llll@{}}
\toprule
\textbf{Family} & \textbf{Axis A} & \textbf{Axis B} & \textbf{Ref.} \\
\midrule
Classical iterative (TV, PWLS-TV)        & in-loop    & hand-crafted & \cite{sidky2008tv,wu2015sparseview} \\
Image-domain denoiser (bilateral, U-Net) & none       & hand-cr. / sup. & \cite{manduca2009bilateral,wagner2022bilateral} \\
Unrolled learned-iterative (ITNet, U-Swin) & in-loop  & supervised & \cite{wurfl2016deep,wuerfl2018,xu2024hybrid,liang2021swinir,liu2021swin,zhang2021transct} \\
Variational network (Hammernik)          & in-loop    & supervised & \cite{hammernik2017deep,hammernik2018vn} \\
Learned primal-dual (LPD)                & in-loop    & supervised & \cite{adler2018lpd} \\
Dual-domain (DD sup., bilateral, N2I)     & single / in-loop & sup. / self-sup. & \cite{wagner2023dualdomain,hendriksen2020noise2inverse} \\
Generative prior (fastdiff $\times4$)     & in-loop    & generative & \cite{shi2026dm4ct,friedrich2024wdm,song2022scorebased,chung2023dps} \\
Per-scene implicit (NAF, R$^2$-Gaussian) & per-scene  & implicit & \cite{zha2022naf,zha2024r2gaussian,liu2025review} \\
Foundation zero-shot (RAM)               & single     & foundation & \cite{terris2025ram} \\
Band decomposition (Wu-2015)             & in-loop    & hand-cr. / sup. & \cite{wu2015sparseview} \\
\textbf{Recombination / param-efficient (ours)} & \textbf{in-loop} & \textbf{hybrid} & --- \\
\bottomrule
\end{tabular}}
\end{table}

\subsection{Datasets and geometry}\label{sec:data}

\textbf{Mayo low-dose CT.} Real helical low-dose data from the 2016 AAPM Low-Dose CT Grand Challenge.\cite{mccollough2017mayo} Rebinning helical to fan-beam via single-slice rebinning\cite{noo1999singleslice} allows direct use of the fan-beam projector. We built and validated this pipeline, and use a fixed train/val/test split (Supplement~S2).

\begin{figure}[t]
\centerline{\includegraphics[width=0.62\linewidth]{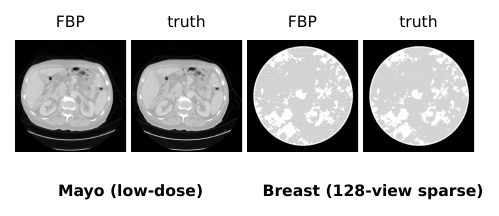}}
\caption{Example slices from the two problems, each showing the FBP input the solvers start from and the ground truth. Left: Mayo low-dose CT, where quantum noise (visible graininess) is the bottleneck. Right: the 128-view sparse-view breast phantom, where undersampling produces streak artifacts. Images are windowed for display, and the FBP is intensity-calibrated to the truth as in the metric.\label{fig:dataex}}
\end{figure}

\textbf{Breast CT Challenge.} 128-view 2-D fan-beam sparse-view data of synthetic breast phantoms.\cite{sidky2022} This data is \textbf{noiseless by design}. The challenge asks whether deep learning can solve the sparse-view inverse problem, so its RMSE floor is exactly zero. We re-partitioned the public 4000-case train pool into train (3600), validation (200), and test (200), with a fixed seed and disjoint cases. To confirm the split, we verified disjointness by image hashing: no test image appears in train or validation (Supplement~S3). Figure~\ref{fig:dataex} shows an example slice from each problem.

\subsection{The calibrated-headroom metric}\label{sec:metric}

We score how far a reconstruction closes the gap between the FBP baseline and the reference. Let $\mathbf{M}$ be a field-of-view mask. For a reconstruction $\hat{\mathbf{x}}$ the headroom is
\begin{equation}\label{eq:hr}
\mathrm{hr}(\hat{\mathbf{x}})=\max\!\Big(0,\ 1-\frac{\mathrm{RMSE}\big(\mathbf{M}\!\odot\!\hat{\mathbf{x}},\,\mathbf{M}\!\odot\!\mathbf{x}\big)}{\mathrm{RMSE}\big(\mathbf{M}\!\odot\!\mathbf{x}_{\mathrm{FBP}},\,\mathbf{M}\!\odot\!\mathbf{x}\big)}\Big).
\end{equation}
Here $\mathbf{M}$ is the circular scan field-of-view mask and $\mathbf{x}$ is the reference: the noiseless phantom image on breast CT, and the high-dose scan on Mayo. $\mathbf{x}_{\mathrm{FBP}}$ is the filtered-back-projection baseline reconstructed from the same measurements the method itself receives, so hr always measures the improvement over doing nothing on that input: the low-dose sinogram on Mayo, the clean sinogram on the noiseless breast board, and the noisy sinogram on the two noisy breast boards. An hr of 0 means no better than that FBP baseline: the $\max(0,\cdot)$ floors any method that fails to beat it. An hr near 1 means near-perfect. The single error hr rewards is masked RMSE against the truth. hr is the only quantity the agent optimises and the only ranking metric; SSIM and PSNR are reported alongside as secondary, human-facing checks (on a batch-wide data range so they are comparable across cases), never optimised or ranked on. We select each method's best iteration on validation, then report its scores once on the held-out test set as the per-case mean $\pm$ standard deviation.

\subsection{The noise-robustness experiment}\label{sec:noiseexp}

We ask whether the noiseless ranking survives a little input noise. For each method we take its noiseless best iteration. We then re-evaluate it on the \textbf{same} 200 breast test cases, but with photon noise added to the sinograms. For a line integral $p$ and incident photon count $I_0$,
\begin{equation}\label{eq:noise}
N\sim\mathrm{Poisson}\big(I_0\,e^{-p}\big),\qquad \hat p=-\log\!\big(\max(N,1)/I_0\big).
\end{equation}
We use $I_0=10^5$ photons. This is a rather high dose. The noise is mild --- about 1--2\,\% at the thickest ray. We apply it to the test inputs only. Note that the ground truth stays clean. Supervised solvers \textbf{load their noiseless-trained weights and no additional training is applied}. Per-scene and classical solvers re-fit on the noisy input, which is their normal inference. The FBP baseline in Eq.~(\ref{eq:hr}) is recomputed from the same noisy data, so hr measures improvement over the noisy FBP. This is exactly the realistic model $\mathbf{A}\mathbf{x}+\boldsymbol{\varepsilon}$ that the noiseless challenge does not study. In a second step, we retrained all models on the noisy data and checked the performance again, to see whether the rankings recover.

\subsection{Statistics}\label{sec:stats}

We begin with Mayo. There the five held-out patients give $n=5$, so the paired $t$-test has low power; we therefore report both the 5\,\% and 1\,\% levels. For Breast CT, the 200 test cases are shared by all methods, so comparisons are paired. With $n=200$ the paired $t$-test is very high-powered: tiny mean gaps reach $p<0.01$. Hence, we report \textbf{effect size} --- how large a difference is, independent of sample size --- not just $p$. We use Cohen's $d_z=\bar d/s_d$, the mean of the paired differences over their standard deviation, and lead with $d_z$ and the raw difference. The contrast between $n=5$ and $n=200$ is discussed in Section~\ref{sec:discussion}.

\section{Results}\label{sec:results}

\subsection{Mayo low-dose CT: a tier, not a winner}\label{sec:tier}

On Mayo the leaderboard yields no single winner, only a small top tier that is statistically indistinguishable. The champion is ITNet at hr $0.376\pm0.089$, with ITNet-v2 (0.374) and U-Swin (0.370) inside one standard error. A paired $t$-test cannot separate ITNet, its variants, and U-Swin at the 5\,\% level, and our compact recombination solver (Section~\ref{sec:compact}; listed as \emph{param-efficient} in the tables and figures) joins this tier at the 1\,\% level ($p=0.02$--$0.027$). The cause is sample size: with five patients ($n=5$), only large, consistent gaps clear significance. The paired $t$-test separates the wider gaps below the tier but not the small, patient-to-patient-noisy differences among the top methods (a two-sided Wilcoxon signed-rank test cannot reach $p<0.05$ at all at $n=5$), so the top of the board is best read as a tie. The absolute image metrics agree: the top six solvers all exceed SSIM 0.97 and sit at the RMSE floor of $5\times10^{-4}$, so SSIM saturates exactly where headroom still separates methods. Below the tier the board spreads cleanly, from the variational network (hr 0.159) and classical total variation (0.096) down through the PWLS-TV and bilateral baselines. Several families (diffusion priors, per-scene implicit methods, the zero-shot foundation model, and the primal-dual unroll as configured for the Mayo geometry) do not clear the low-dose FBP baseline and sit at hr 0 even though their SSIM stays high (0.82--0.95): SSIM rewards the smoothness that hr, measured against FBP, does not credit. The full board, per-patient headroom, and pairwise significance matrix are in Table~\ref{tab:mayo} and Supplement~S4.

\begin{table}[t]
\caption{Mayo low-dose CT test board, ranked solvers ($n=5$ test patients). hr is the calibrated headroom (Eq.~\ref{eq:hr}) and the ranking metric; SSIM and RMSE use a batch-wide data range (RMSE in line-integral units). Params are trainable-parameter counts (M = millions, otherwise a raw count; dd = dual-domain, sup = supervised). The 11 solvers that did not clear the FBP floor (hr $\le 0$) are omitted here; the full 26-solver board is in Supplement~S4.\label{tab:mayo}}
\scriptsize
\setlength{\tabcolsep}{4pt}
\centering
\begin{tabular}{@{}rlrrrr@{}}
\toprule
\textbf{Rank} & \textbf{Solver} & \textbf{hr} & \textbf{SSIM} & \textbf{RMSE} & \textbf{Params} \\
\midrule
1  & itnet             & 0.3756 & 0.9790 & 0.0005 & 0.233\,M \\
2  & itnet-v2          & 0.3735 & 0.9784 & 0.0005 & 0.233\,M \\
3  & uswin             & 0.3700 & 0.9770 & 0.0005 & 3.954\,M \\
4  & dd-supervised     & 0.3607 & 0.9758 & 0.0005 & 0.466\,M \\
5  & param-efficient   & 0.3241 & 0.9727 & 0.0005 & 969 \\
6  & itnet-v3          & 0.3066 & 0.9707 & 0.0006 & 3.699\,M \\
7  & hammernik-vn      & 0.1591 & 0.9520 & 0.0007 & 0.005\,M \\
8  & tv-iterative      & 0.0959 & 0.9642 & 0.0007 & 0 \\
9  & hammernik-2017    & 0.0843 & 0.9358 & 0.0007 & 0.012\,M \\
10 & manhart-pwls-tv   & 0.0624 & 0.9649 & 0.0007 & 0 \\
11 & wu-2015-trainable & 0.0207 & 0.8923 & 0.0008 & 8 \\
12 & dd-n2i            & 0.0176 & 0.9590 & 0.0008 & 0.466\,M \\
13 & manduca-bilateral & 0.0159 & 0.9561 & 0.0009 & 7 \\
14 & tv-iterative-sup  & 0.0109 & 0.9366 & 0.0008 & 2 \\
15 & dd-bilateral-sup  & 0.0040 & 0.9541 & 0.0009 & 12 \\
\bottomrule
\end{tabular}
\end{table}

\subsection{A meta-result: instructing is far cheaper than coding}\label{sec:effort}

One outcome deserves emphasis on its own, because it bears directly on using agents for research: the effort accounting. Every line of code in this study, the geometry pipeline, all solvers, and the benchmark harness, was written by the agent. The human did not hand-code but instructed. What varied, by a wide margin, was how much instruction each part needed. The helical-to-fan \emph{geometry pipeline} was by far the hardest: bringing the differentiable rebinning and its calibration to a physically correct FBP baseline took many supervised iterations and was the dominant human cost of the study (Supplement~S2). \emph{Implementing and benchmarking the 26 solvers} needed far less. The agent onboarded 13 of them across two days, and the three-week benchmark campaign was overwhelmingly automated, most of its commits changing only results and not code, while the agent ran $\sim$1{,}100 iterations for $\sim$134~GPU-hours of compute, largely in parallel. Costing each source-changing commit at an estimated 7.5 minutes (Supplement~S2), the human's hands-on share of the solver and benchmarking work was on the order of hours. Instructing an agent is far cheaper than coding by hand, though not uniformly, since the hard, physics-heavy pipeline still demanded the most human supervision while the methods layered on top cost only hours.

\subsection{Noiseless Breast CT Challenge: the numbers are not a leak}\label{sec:noleak}

On the noiseless Breast CT Challenge the headroom values are high. The champion is the supervised dual-domain denoiser at hr 0.89, with the ITNet family (0.87--0.89) and U-Swin (0.86) just behind; the top five solvers all exceed 0.85 (Table~\ref{tab:reversal}, left board). This is a property of the data, not a flaw and not a leak. The Breast CT Challenge data is noiseless by design, so its RMSE floor is exactly zero and near-perfect recovery is possible in principle. Because such numbers invite suspicion, we ruled out train/test leakage two ways. By construction, the public pool is partitioned by index under a fixed seed, so no case can appear in more than one split; and by mechanism, the training loader only ever reads the train split, with the test redirect firing solely at evaluation, so no model sees a test image while training. Hence the high numbers reflect an easy, well-posed, incompleteness-limited problem, not memorization. This also means the breast and Mayo hr scales are not comparable. Breast is limited by missing views (128 of $\sim$1000), Mayo by real quantum noise, so a breast hr of 0.89 and a Mayo hr of 0.38 describe different bottlenecks, not different amounts of the same thing (Supplement~S3).

\subsection{A compact, problem-dependent solver}\label{sec:compact}

\begin{figure}[t]
\centerline{\includegraphics[width=0.62\linewidth]{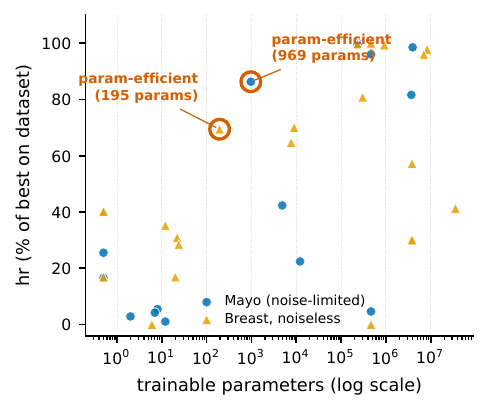}}
\caption{Accuracy versus parameter count for both datasets (Mayo and breast noiseless). To overlay the two different hr scales, each solver's hr is normalized to the percent of the best hr reached on its own board. The agent's compact recombination solver is ringed: 969 parameters on Mayo, 195 on breast, i.e.\ 0.4\,\% and 0.04\,\% of the respective champion's count; on Mayo it ties the top tier at the 1\,\% level.\label{fig:params}}
\end{figure}

Beyond ranking the 26 established methods, the agent recombined their parts into a compact solver, and the best recombination is problem-dependent. On Mayo, a noise-limited problem, the compact optimum is a small multi-scale bilateral denoiser of about 970 parameters (hr 0.324) that joins the top tier at the 1\,\% level: where the bottleneck is noise, a tiny denoiser suffices. On breast, a sparse-view problem, the optimum is entirely different: a filtered data-consistency step ($\mathbf{D}=\mathrm{FBP}$ applied to the residual $\mathbf{A}\hat{\mathbf{x}}-\mathbf{g}$), a learned cross-step combination in the primal-dual style (Eq.~\ref{eq:lpd}), and a small multi-scale bilateral output filter. The climb to it was single-knob iterations, each earned against the frozen metric: a pure image-domain denoiser, however configured, capped at hr $\approx0.26$; replacing the adjoint by the \emph{filtered} gradient broke that ceiling with only 13 parameters (hr 0.43); a learned cross-step combination added headroom (183 parameters, hr 0.50); and a three-scale bilateral output filter completed the design (195 parameters). The final breast solver reaches hr $0.6212\pm0.0076$ with 195 parameters, three orders of magnitude fewer than the champion's 0.47\,M (hr 0.895), and with the tightest per-case standard deviation of any solver on the board. The two compact solutions do not transfer: on a sparse-view problem data-consistency is the differentiator, on a noise-limited problem it is not. This is the practical face of the known-operator principle: embedding the exact forward operator cannot raise, and usually lowers, the maximum error bound, and it lets a few hundred parameters do the work of millions.\cite{maier2019knownop} The agent re-derived each optimum from its own runs; steered toward the Mayo answer on breast it did not transfer, because the breast evidence pointed elsewhere (Supplement~S5). Figure~\ref{fig:params} places both compact solvers against parameter count.

\subsection{A little noise inverts the ranking}\label{sec:reversal}

We now perturb the inputs and re-score, with no retraining. We add mild Poisson noise to the 200 breast test sinograms at $I_0=10^5$ photons (about 1--2\,\% at the thickest ray), keep the ground truth clean, reload each model's noiseless-trained weights, and recompute the FBP baseline from the noisy data so that hr measures improvement over the \emph{noisy} FBP. Nothing else changes. The clean ranking does not predict the noisy one. Across the solvers ranked on clean data, the two orderings are essentially uncorrelated (Spearman $\rho\approx0.04$; the clean ranking explains under 1\,\% of the variance in noisy rank). The noiseless champion drops to the zero-headroom group at the bottom, and none of the five best solvers on clean data survive in the noisy top five; they are displaced by physics-regularized and hand-crafted-smoothing methods. Table~\ref{tab:reversal} shows the two boards side by side. The headline entries:

\begin{itemize}
\item \textbf{learned-primal-dual}\cite{adler2018lpd} rises from hr 0.72 (rank 6) to \textbf{0.93} (rank 1), with noisy SSIM 0.99. It keeps the forward operator in the loop, so it re-solves the noisy measurements rather than denoising a fixed image.
\item \textbf{manduca-bilateral}, a 22-parameter hand-crafted smoothing prior,\cite{manduca2009bilateral} rises from 0.28 to \textbf{0.84} (rank 2). It has nothing to overfit, so smoothing is noise-robust by construction. The self-supervised dual-domain bilateral (dd-bilateral-n2i)\cite{hendriksen2020noise2inverse} rises even more sharply, from an essentially failed 0.0001 on clean data to \textbf{0.79} (rank 3).
\item \textbf{dual-domain-supervised}, the noiseless champion (0.89, rank 1),\cite{wagner2023dualdomain} \textbf{collapses to hr 0.00} and out of the ranking, with SSIM 0.35. It is a pure supervised image-domain denoiser trained only on clean FBP, so a distribution it never saw breaks it.
\item The supervised unrolled family (ITNet v1/v2/v3, U-Swin) drops from 0.86--0.89 into the middle of the board (0.55--0.70). Our compact solver holds relatively well (0.62 $\to$ 0.52, rank 14).
\item Tellingly, the unconstrained wavelet diffusion prior (fastdiff-wdm-wav-unc), a deliberate negative control that \emph{failed} on clean data (hr 0, below the FBP floor), is noise-robust and ranks 9th (hr 0.64, SSIM 0.95). Its generative prior never fit the clean forward model, so it degrades gracefully, and the zero-shot foundation model (RAM)\cite{terris2025ram} rises for the same reason (rank 8).
\item Per-scene neural-field and Gaussian-splatting methods (NAF, R$^2$-Gaussian)\cite{zha2022naf,zha2024r2gaussian} do not complete under the 20-minute budget on either board and are reported as did not finish (DNF).
\end{itemize}

\begin{figure}[t]
\centerline{\includegraphics[width=0.62\linewidth]{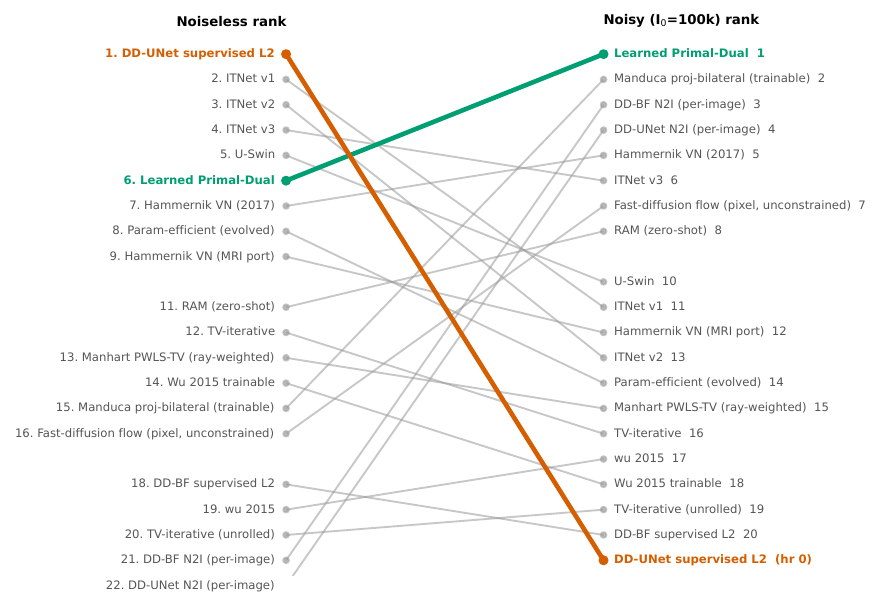}}
\caption{The reversal. Paired rank plot from the noiseless breast board to the noisy breast board ($I_0=10^5$, same trained models, same 200 test cases, no retraining), with an arrow per method. The supervised image-domain champion collapses from rank~1 to last; the learned primal-dual method rises from mid-pack to champion; hand-crafted smoothing and in-loop physics priors rise. A small, realistic noise perturbation reorders almost the whole leaderboard.\label{fig:reversal}}
\end{figure}

\begin{figure}[t]
\centerline{\includegraphics[width=0.80\linewidth]{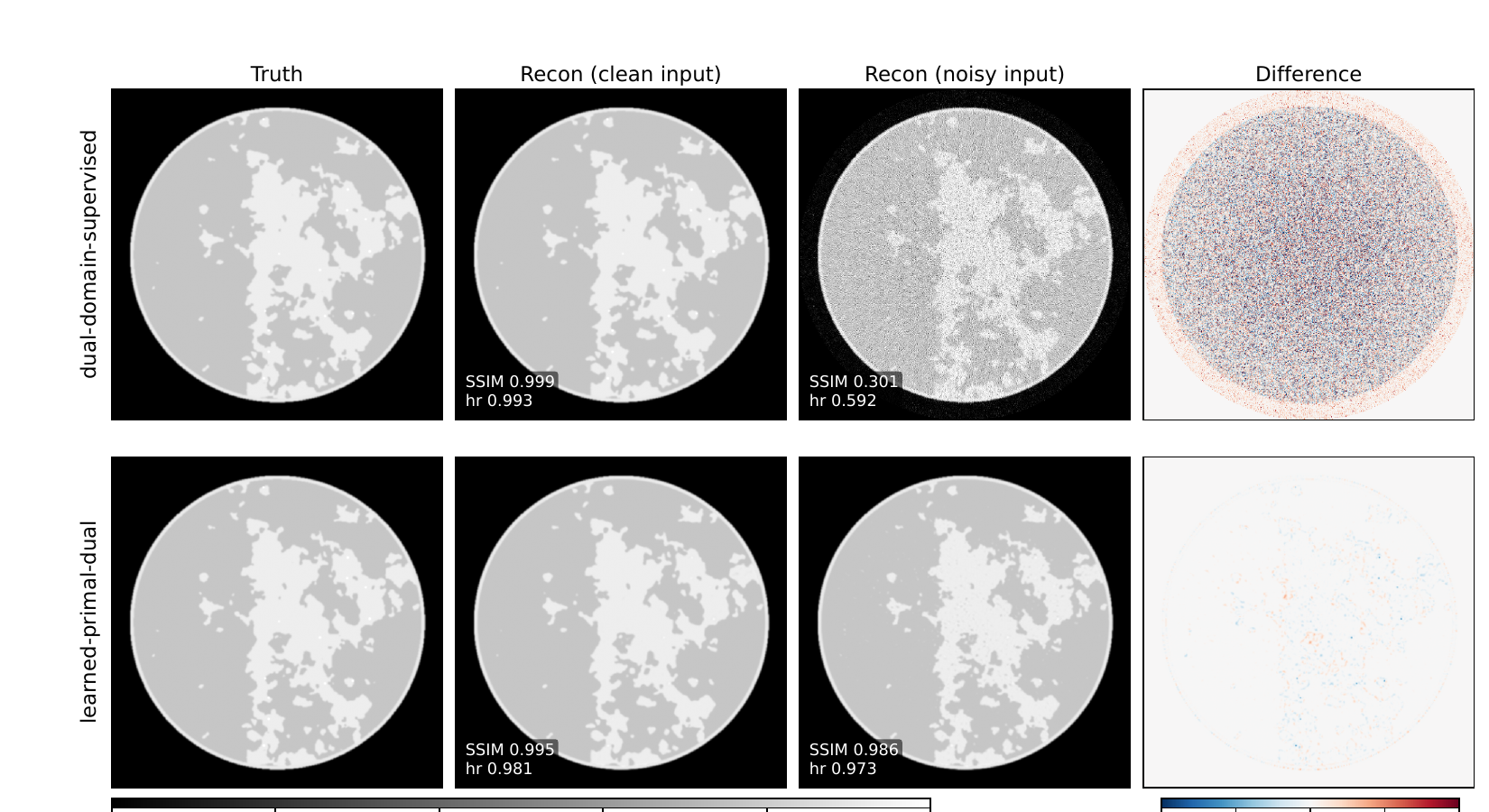}}
\caption{Example reconstructions behind the reversal, for the two methods at opposite ends of it. Rows: the noiseless champion \emph{dual-domain-supervised} (a supervised image-domain denoiser) versus \emph{learned-primal-dual} (in-loop physics). Columns: ground truth, reconstruction from the clean input, reconstruction from the noisy input ($I_0=10^5$, same trained model, no retraining), and the difference to truth. Per-case SSIM and hr are annotated on each reconstruction; these are single-case values and differ from the 200-case board means in Table~\ref{tab:reversal}. On clean inputs both are near-perfect. Under mild input noise the supervised denoiser collapses (single-case SSIM $0.999\!\to\!0.301$), spraying structured error across the field of view, while the physics-in-the-loop primal-dual method holds (SSIM $0.995\!\to\!0.986$). This is the pixel-level face of the ranking reshuffle in Figure~\ref{fig:reversal} and Table~\ref{tab:reversal}.\label{fig:panels}}
\end{figure}

The absolute SSIM and PSNR columns confirm the reordering is genuine, not an artifact of the changing FBP baseline: learned-primal-dual reaches SSIM 0.99 on the noisy data while dual-domain-supervised falls to 0.35 (Supplement~S6). Because every method sees the identical noisy input and shares the same noisy FBP baseline, the comparison is fair within the noisy board. Figure~\ref{fig:reversal} shows the full reordering as a paired rank plot, and Figure~\ref{fig:panels} shows its pixel-level face for the two methods at opposite ends.

\subsection{The framework explains the reversal}\label{sec:explain}

The reversal is not a surprise once we read it off the two axes of Table~\ref{tab:taxonomy}. Brittleness concentrates in one corner: supervised priors with weak physics engagement, i.e.\ image-domain maps trained on clean FBP. Dual-domain-supervised is the extreme case: a learned image-to-image map with no forward operator at inference, so a small input shift it never saw during training moves it off its learned manifold and the reconstruction collapses. The constrained diffusion variants, which also lean on a clean-data fit, fall the same way. Robustness concentrates where the prior is not tuned to the clean distribution, and two disjoint mechanisms reach it. The first is in-loop physics: learned-primal-dual and the variational networks keep the forward operator $\mathbf{A}$ in the loop, so they re-solve the noisy measurements instead of denoising a fixed image, and their ranking rises or holds. The second is a prior that never fit the clean data at all: hand-crafted smoothing (bilateral, total variation) has nothing to overfit, and even the generative and foundation priors that failed on clean data degrade gracefully because they never matched the clean forward model. The framework thus turns the reversal from a curiosity into a prediction: a method's position on the physics-engagement and prior-source axes tells us, before any noise is added, whether it is likely to be brittle or robust. The lesson for benchmarking follows directly: a leaderboard measured on a single idealized distribution rewards exactly the methods most likely to fail under a small, realistic perturbation.

\subsection{Retraining on the noise restores the ranking}\label{sec:retrain}

The reversal above is a \emph{no-retrain} result: it measures how clean-optimal models transfer to an unseen noise level --- the realistic deployment question, where a model shipped after training on clean data later meets noisier inputs. The converse question is also worth asking: if the noise is known at training time, does the clean ranking return? We retrained each trainable solver's clean-best configuration --- unchanged architecture and hyper-parameters --- from scratch on the same breast training set with matched Poisson noise added ($I_0=10^5$), selected the checkpoint on the noisy validation split, and scored it on the noisy test set. Classical, per-scene, and zero-shot solvers carry no trainable weights and are unchanged. We retrained only the frozen clean-best configuration and did not re-search hyper-parameters, so these numbers are a lower bound on what matched-noise training could reach.

The clean ranking largely returns. Its Spearman correlation with the noisy board rises from $\rho\approx0.04$ without retraining (under 1\,\% of the variance) to $\rho\approx0.61$ with it (37\,\%). Table~\ref{tab:retrain} shows the recovery, plotted as a three-board rank flow in Supplement~S6. The noiseless champion, dual-domain-supervised, which had collapsed to hr $0.00$, retrains to \textbf{hr $0.958$} and reclaims rank~1, above its own noiseless score, because hr is now measured against the worse noisy FBP baseline. Its collapse was therefore a train/test distribution mismatch, not a limit of the architecture, exactly as the framework predicts. The whole supervised unrolled family (ITNet v1/v2/v3, U-Swin) returns to the top tier (hr $\approx0.94$). Learned-primal-dual, already robust because it keeps the forward operator in the loop, barely moves ($0.935\!\to\!0.936$): it had nothing to regain.

Two groups do not recover. The two \emph{constrained} fast-diffusion variants stay at hr $0$; their hard fit to the clean forward model does not survive noisy training. And a few solvers dip slightly (Hammernik-2017, $0.70\!\to\!0.61$) because we retrained the fixed clean-best configuration rather than re-searching the noisy optimum. The lesson for benchmarking, though, deepens rather than settles. Noise proved the most tractable confounder, since one retraining pass repaired it, but two things stand out. Adding it to the breast training data did not restore the breast problem. It built a third, blended challenge, sparse-view breast under Mayo-like noise. And noise is only one of an open-ended, evolving set of shifts (beam hardening, scatter, the anatomy, the disease and biology under study). The recovery is real, but it only reinforces that a challenge isolating a single factor cannot certify a method's \emph{generality} (Section~\ref{sec:findings}).

\begin{table}[t]
\caption{Retraining on the noise restores the ranking. Per-solver headroom (hr) on the three breast boards: noiseless; noisy with \emph{no} retraining ($I_0=10^5$, frozen clean weights); and noisy after \emph{retraining} each solver's clean-best configuration on matched-noise training data. Rows are ordered by the retrained rank. ``---'' = did not clear the hr floor (hr $\le0$) or did not finish. $^\dagger$ marks a classical / per-scene / zero-shot solver with no trainable weights, carried over unchanged. The clean-favored supervised methods, which collapse without retraining, return to the top: Spearman correlation with the noiseless ranking rises from $\rho\approx0.04$ (no retrain) to $\rho\approx0.61$ (retrained).\label{tab:retrain}}
\scriptsize
\setlength{\tabcolsep}{4pt}
\centering
\begin{tabular}{@{}rlrrr@{}}
\toprule
\textbf{Rank} & \textbf{Solver} & \textbf{Noiseless} & \textbf{Noisy} & \textbf{Noisy} \\
              &                 & \textbf{hr}        & \textbf{no-retr.} & \textbf{retrain} \\
\midrule
 1 & dd-supervised & 0.895 & 0.000 & 0.958 \\
 2 & itnet-v2 & 0.889 & 0.548 & 0.939 \\
 3 & itnet-v3 & 0.875 & 0.701 & 0.938 \\
 4 & itnet & 0.893 & 0.578 & 0.938 \\
 5 & learned-primal-dual & 0.723 & 0.935 & 0.936 \\
 6 & uswin & 0.859 & 0.592 & 0.936 \\
 7 & hammernik-vn & 0.579 & 0.553 & 0.903 \\
 8 & param-efficient & 0.621 & 0.515 & 0.851 \\
 9 & manduca-bilateral$^\dagger$ & 0.276 & 0.842 & 0.842 \\
10 & dd-bilateral-sup & 0.255 & 0.100 & 0.821 \\
11 & dd-bilateral-n2i & 0.000 & 0.793 & 0.789 \\
12 & tv-iterative-sup & 0.151 & 0.107 & 0.773 \\
13 & dd-n2i & 0.000 & 0.725 & 0.725 \\
14 & fastdiff-flow-px-unc & 0.269 & 0.662 & 0.662 \\
15 & ram-zeroshot$^\dagger$ & 0.369 & 0.638 & 0.638 \\
16 & fastdiff-wdm-wav-unc & 0.000 & 0.635 & 0.635 \\
17 & hammernik-2017 & 0.627 & 0.702 & 0.610 \\
18 & manhart-pwls-tv$^\dagger$ & 0.360 & 0.507 & 0.507 \\
19 & tv-iterative$^\dagger$ & 0.360 & 0.507 & 0.507 \\
20 & wu-2015$^\dagger$ & 0.152 & 0.370 & 0.370 \\
21 & wu-2015-trainable & 0.316 & 0.259 & 0.268 \\
-- & fastdiff-wdm-wav-con & 0.269 & 0.000 & 0.000 \\
-- & fastdiff-flow-px-con & 0.512 & 0.000 & 0.000 \\
-- & naf$^\dagger$ & --- & --- & --- \\
-- & r2gaussian$^\dagger$ & --- & --- & --- \\
\bottomrule
\end{tabular}
\end{table}

\begin{table}[t]
\caption{The two breast boards side by side, for the same trained models on the same 200 test cases with \emph{no retraining}. Left: noiseless board. Right: noisy board ($I_0=10^5$). The ranking nearly inverts. hr is the calibrated headroom (Eq.~\ref{eq:hr}) and the ranking metric; SSIM and RMSE use a batch-wide data range (RMSE in line-integral units). ``--'' marks a solver that did not clear the ranking floor (hr $\le 0$) or did not finish (---). Params (M suffix = millions, otherwise a raw count) and solver keys as in Table~\ref{tab:mayo}.\label{tab:reversal}}
\scriptsize
\setlength{\tabcolsep}{3pt}
\begin{minipage}[t]{0.49\textwidth}
\centering
\textbf{Noiseless board}\\[2pt]
\begin{tabular}{@{}rlrrrr@{}}
\toprule
\textbf{Rank} & \textbf{Solver} & \textbf{hr} & \textbf{SSIM} & \textbf{RMSE} & \textbf{Params} \\
\midrule
1  & dd-supervised        & 0.8948 & 0.9992 & 0.0005 & 0.466\,M \\
2  & itnet                & 0.8926 & 0.9991 & 0.0006 & 0.233\,M \\
3  & itnet-v2             & 0.8893 & 0.9991 & 0.0006 & 0.928\,M \\
4  & itnet-v3             & 0.8749 & 0.9989 & 0.0007 & 8.318\,M \\
5  & uswin                & 0.8586 & 0.9986 & 0.0007 & 7.025\,M \\
6  & learned-primal-dual  & 0.7233 & 0.9962 & 0.0014 & 0.304\,M \\
7  & hammernik-2017       & 0.6265 & 0.9902 & 0.0019 & 0.009\,M \\
8  & param-efficient      & 0.6212 & 0.9912 & 0.0020 & 195 \\
9  & hammernik-vn         & 0.5787 & 0.9865 & 0.0022 & 0.008\,M \\
10 & fastdiff-flow-px-con & 0.5119 & 0.9792 & 0.0025 & 3.823\,M \\
11 & ram-zeroshot         & 0.3693 & 0.9787 & 0.0033 & 35.619\,M \\
12 & tv-iterative         & 0.3601 & 0.9918 & 0.0033 & 0 \\
13 & manhart-pwls-tv      & 0.3601 & 0.9918 & 0.0033 & 0 \\
14 & wu-2015-trainable    & 0.3156 & 0.9537 & 0.0035 & 12 \\
15 & manduca-bilateral    & 0.2763 & 0.9830 & 0.0037 & 22 \\
16 & fastdiff-flow-px-unc & 0.2693 & 0.9776 & 0.0038 & 3.823\,M \\
17 & fastdiff-wdm-wav-con & 0.2693 & 0.9340 & 0.0038 & 3.825\,M \\
18 & dd-bilateral-sup     & 0.2546 & 0.9848 & 0.0039 & 24 \\
19 & wu-2015              & 0.1517 & 0.9611 & 0.0044 & 0 \\
20 & tv-iterative-sup     & 0.1513 & 0.9613 & 0.0044 & 20 \\
21 & dd-bilateral-n2i     & 0.0001 & 0.9583 & 0.0061 & 6 \\
22 & dd-n2i               & 0.0001 & 0.9461 & 0.0062 & 0.466\,M \\
-- & fastdiff-wdm-wav-unc & 0.0000 & 0.9448 & 0.0112 & 3.825\,M \\
-- & naf                  & ---    & ---    & ---    & 33.563\,M \\
-- & r2gaussian           & ---    & ---    & ---    & 0.006\,M \\
\bottomrule
\end{tabular}
\end{minipage}\hfill
\begin{minipage}[t]{0.49\textwidth}
\centering
\textbf{Noisy board} ($I_0=10^5$)\\[2pt]
\begin{tabular}{@{}rlrrrr@{}}
\toprule
\textbf{Rank} & \textbf{Solver} & \textbf{hr} & \textbf{SSIM} & \textbf{RMSE} & \textbf{Params} \\
\midrule
1  & learned-primal-dual  & 0.9349 & 0.9905 & 0.0020 & 0.304\,M \\
2  & manduca-bilateral    & 0.8420 & 0.9537 & 0.0049 & 22 \\
3  & dd-bilateral-n2i     & 0.7927 & 0.9532 & 0.0064 & 6 \\
4  & dd-n2i               & 0.7252 & 0.8260 & 0.0084 & 0.466\,M \\
5  & hammernik-2017       & 0.7016 & 0.7178 & 0.0092 & 0.009\,M \\
6  & itnet-v3             & 0.7007 & 0.7564 & 0.0092 & 8.318\,M \\
7  & fastdiff-flow-px-unc & 0.6620 & 0.7058 & 0.0104 & 3.823\,M \\
8  & ram-zeroshot         & 0.6382 & 0.6578 & 0.0111 & 35.619\,M \\
9  & fastdiff-wdm-wav-unc & 0.6350 & 0.9450 & 0.0112 & 3.825\,M \\
10 & uswin                & 0.5915 & 0.6408 & 0.0125 & 7.025\,M \\
11 & itnet                & 0.5784 & 0.6295 & 0.0129 & 0.233\,M \\
12 & hammernik-vn         & 0.5529 & 0.5962 & 0.0137 & 0.008\,M \\
13 & itnet-v2             & 0.5480 & 0.6201 & 0.0139 & 0.928\,M \\
14 & param-efficient      & 0.5153 & 0.5771 & 0.0149 & 195 \\
15 & manhart-pwls-tv      & 0.5068 & 0.6164 & 0.0151 & 0 \\
16 & tv-iterative         & 0.5068 & 0.6164 & 0.0151 & 0 \\
17 & wu-2015              & 0.3700 & 0.5244 & 0.0193 & 0 \\
18 & wu-2015-trainable    & 0.2591 & 0.4400 & 0.0227 & 12 \\
19 & tv-iterative-sup     & 0.1071 & 0.4567 & 0.0274 & 20 \\
20 & dd-bilateral-sup     & 0.1003 & 0.4524 & 0.0276 & 24 \\
-- & dd-supervised        & 0.0000 & 0.3483 & 0.0338 & 0.466\,M \\
-- & fastdiff-wdm-wav-con & 0.0000 & 0.3104 & 0.0477 & 3.825\,M \\
-- & fastdiff-flow-px-con & 0.0000 & 0.2769 & 0.0583 & 3.823\,M \\
-- & naf                  & ---    & ---    & ---    & 33.563\,M \\
-- & r2gaussian           & ---    & ---    & ---    & 0.006\,M \\
\bottomrule
\end{tabular}
\end{minipage}
\end{table}

\section{Discussion}\label{sec:discussion}

\subsection{What the agent did well, and what it did not}\label{sec:scorecard}

We give a scorecard. It is the core of this Discussion.

\textbf{The agent is strong in four ways.} First, it is fast from paper to code: it turns a method description into a working solver quickly. The version-control history quantifies this. The agent onboarded 13 of the 26 established solvers across only two days (15--16 May 2026) and the rest in small batches; from a published method to a running, benchmarked solver was a fraction of a day each (per-solver dates in Supplement~S2). This is the main speed-up, and it is why 26 methods across two datasets was feasible. In fact, it is also strong at hyper-parameter optimization. It respects a fixed compute budget, which is hard for humans and easy for the agent. And it scales evaluation: running many benchmarks in parallel is its strongest practical use.

\textbf{The agent is weak in several ways.} It does not invent new methods. Rather, it implements, tunes, and recombines known ones, and proposed no new reconstruction principle. Yet, it had to be \emph{forced} to mix and match. The idea of combining proven pieces into one compact solver was a human idea, on both datasets. The agent executed it well but did not originate it. Left alone, it converges early and stops exploring. The compact-solver search shows this directly: the agent's own first 20 iterations plateaued at hr $\approx0.50$, and the decisive gains (the break to hr 0.62) came only after a human began suggesting untried directions from iteration~20 on (Supplement~S5). Its CT-image vision is unreliable: it keeps missing very clear artifacts in slices and sinograms, so numbers had to be the source of truth. Long tasks must be decomposed into sub-tasks; otherwise even a one-million-token context is used up quickly. And it overfits to the task and metric. Of course, this last point is probably not only an agent problem, since human researchers overfit to benchmarks too.

\textbf{The agent could not steer on the test set.} By design, held-out test performance never entered the search. The agent's feedback each iteration was the validation metric and the validation result images, the same images published on the live dashboard, and every model selection used validation alone. The challenge data resided only on the compute cluster, read at run time by each solver's data loader, and the test set was reconstructed and scored a single time, after the search had ended, to build the final leaderboards. The agent thus never observed a test score while it worked, so the reported held-out results cannot have been selected for. Split integrity is audited in Supplement~S3.

\textbf{The division of labor is clear.} On one side, the agent is a tireless, budget-respecting, self-modifying executor. On the other, the human remains the strategist and auditor, forcing breadth, supplying the recombination idea, redirecting across problems, decomposing the work, and checking for padding and provenance.

\subsection{Two scientific findings}\label{sec:findings}

First, the best compact architecture is problem-dependent. The Mayo solution is a denoiser, whereas the breast solution is a filtered-data-consistency and primal-dual unroll. Again, the agent re-derived each; it did not transfer one to the other. This supports the known-operator view: the right inductive bias (the assumptions an architecture builds in before it sees any data) depends on whether the problem is noise-limited or incompleteness-limited.

Second, the clean-data ranking does not predict robustness to noise the models never trained on, but that brittleness is largely trainable away. A mild perturbation reordered almost the whole breast board and sent the clean winners to the bottom. Retraining those same models on matched noise then brought the ranking back (Spearman $\rho$ from $0.04$ to $0.61$) and returned the collapsed champion to first (Section~\ref{sec:retrain}). The collapse is therefore a train/test distribution mismatch, not a fixed deficit of the architecture. But the deeper lesson is about challenge \emph{design}, and it cuts against a first reading of our own result. Noise was the easy case, repaired by one retraining pass. Real reconstruction faces an open-ended set of shifts (beam hardening, scatter, the anatomy in the field of view, the disease and biology under study) that cannot be enumerated and does not hold still, because new pathologies and treatments introduce new shifts at any time. And patching a factor in changes the problem: by adding noise to the breast training data we did not recover the breast challenge but constructed a third, blended one, sparse-view breast under Mayo-like noise. Generality is a claim about the whole space of co-occurring factors, so a benchmark that varies a single factor, even reported under several conditions, cannot certify it. The productive response is not to append confounders one at a time, an endless and always-incomplete list, but to design challenges that deliberately span a broad spectrum of realistic factors at once. Significance also depends on sample size: at $n=5$ (Mayo) the top methods tie, and at $n=200$ (breast) everything separates. As such, $p$-values are not comparable across datasets, and effect size should lead.

\subsection{Limitations}\label{sec:limits}

We used a single agent and a single frozen metric under a fixed short per-iteration compute budget. That single scalar metric limits the search itself, not only the benchmark: with headroom as the sole source of truth the agent optimises greedily and exploits rather than explores. Both compact solvers plateaued until a human suggested untried directions after iteration~20 (Section~\ref{sec:compact}). Better exploration strategies for the loop are open work: not letting the context fill with the agent's own earlier solutions, or automating the human's role with a critic subagent or a multi-agent ensemble that perturbs the greedy descent. Exploration also depends on the model: in our companion agentic MR work\cite{zaiss2026agent4mr} the model family used here optimised the frozen metric relentlessly but rarely tried anything genuinely new, while other frontier models explored markedly more. Periodically delegating planning to a second, more exploratory model is a cheap thing to try. The geometry and data-engineering bottleneck also remains; the agent did not remove it. The agent-written solvers reproduce each method's design but may differ in detail from its official implementation. Our breast split is an internal partition of the public pool, not the withheld official test set, and the Mayo comparison rests on five test patients ($n=5$), limiting statistical power. Some compact solutions are seed-fragile, and two per-scene methods did not complete under the budget. The noise study covers two conditions, transfer to an unseen noise level and matched-noise retraining (Section~\ref{sec:retrain}), but at a single dose ($I_0=10^5$) rather than a sweep, and the retrained board re-trained only each solver's frozen clean-best configuration rather than re-searching the noisy optimum, so those scores are a lower bound. The effect of matched-noise training on clean-trained networks is documented independently.\cite{huang2018some} We leave a full dose sweep and other perturbation probes, such as physically implausible noise, to future work. Finally, we report no clinical or diagnostic endpoint.

\section{Conclusions}\label{sec:conclusions}

An LLM agent can implement, tune, and benchmark 26 CT reconstruction methods under one frozen metric and a fixed compute budget. The outcome of such a benchmark is a small tier of indistinguishable top methods, not one winner. Still, the agent does the labor; it does not replace the human strategist.

The more important message is about benchmarks. We used the agent to scale evaluation, and that scale exposed a problem. Our ideal-data leaderboard did not predict robustness: a small, realistic perturbation reordered almost the whole board and collapsed the methods that won on clean data. That collapse is a distribution mismatch, not a verdict on the methods, and retraining on matched noise restores most of the ranking. But noise is the easy case. Beam hardening, scatter, the anatomy, and the disease and biology under study are further shifts. The list is open-ended and grows as pathologies and treatments change, and adding even one factor to training changes the problem itself, as our retrained board (effectively a blend of the breast and Mayo challenges) shows. No benchmark that isolates a single factor can therefore certify that a method \emph{generalizes}. The field should invest less in challenges that vary one factor at a time and more in challenges that model a broad spectrum of co-occurring, clinically relevant factors together, and re-run them as those factors evolve. Agentic autoresearch is what makes that affordable. That is its real payoff for the field.

\subsection*{Acknowledgments}
We thank the organizers of the AAPM Low-Dose CT Grand Challenge (Mayo Clinic), Emil Sidky, and the organizers of the Breast CT Challenge for making their data available.

\subsection*{Conflicts of Interest}
The authors declare no relevant conflicts of interest.

\subsection*{Data and Code Availability}
The code --- all 26 solver implementations, the compact recombination, the helical-to-fan rebinning pipeline, and the evaluation scripts --- is open source at \url{https://github.com/akmaier/Agent4CT}, with the leaderboards and per-iteration provenance on the live dashboard at \url{https://akmaier.github.io/Agent4CT}. The challenge datasets are available from their original providers under their data-use terms and are not redistributed here: the Mayo low-dose CT data are the de-identified public AAPM Low-Dose CT Grand Challenge / TCIA collection, and the Breast CT Challenge data are synthetic phantoms. No new patient data were collected and institutional review board approval was not required; we comply with all challenge data-use agreements.


\bibliographystyle{unsrtnat}
\bibliography{refs}
\end{document}